\documentclass{aastex62}
\pdfoutput=1
\shorttitle{Barometric Pressure Correction to Gamma-ray Observations}
\shortauthors{Datar et al.}

\begin{document}

\title{Barometric Pressure Correction to Gamma-ray Observations and its Energy Dependence} 

\correspondingauthor{Gauri Datar}
\email{datar.gouri@gmail.com, gdatar16@iigs.iigm.res.in}

\author[0000-0003-4971-6010]{Gauri Datar}
\affil{Indian Institute of Geomagnetism, Navi Mumbai, 410218, India}

\author{Geeta Vichare}
\affil{Indian Institute of Geomagnetism, Navi Mumbai, 410218, India}

\author{Selvaraj Chelliah}
\affil{Equatorial Geophysical Research Laboratory (IIG), Tirunelveli, 627011, India} 

\begin{abstract}

Cosmic rays (CRs) have been studied extensively in the last century to understand the processes in the universe as well as in the solar system.  The CR studies around the world continue even today using the latest technologies. In today's satellite era, although many observations are made from space, CR observations from the ground are still viewed as an important tool. These observations, however, mostly detect the secondary cosmic rays (SCRs) produced via nuclear spallation processes during the interactions of the primary CR with the atmospheric nuclei. Neutron, muon, and gamma particles are the major components of SCRs as observed from the ground.  It is well known that the atmospheric pressure plays an important role in the SCR flux observed on the ground. Barometric pressure correction is standard practice for neutron monitor (NM) data. However, no such correction is applied to gamma-ray, being massless. But the pressure affects the particles such as $e^{\pm}$, $\mu^{\pm}$, which produce gamma rays in the cascade. Subsequently, the indirect pressure dependence of the gamma-ray flux can be anticipated. 

We examine this aspect in detail by studying the gamma-ray counts detected by the NaI (Tl) detector. The present study confirms that there is no correlation between the atmospheric pressure and the total gamma-ray counts collected from all energies. However, the scenario differs when the gamma-ray fluxes of different energies are investigated separately. The gamma rays of energy below $\sim$3 MeV are primarily due to the radioactivity originating from the ground, whereas gamma rays above 3 MeV are mainly produced in the CR cascade. It is observed that the counts of energy above 3 MeV are well anti-correlated with the atmospheric pressure and hence need to be corrected. It is demonstrated that applying the barometric correction formula successfully removes the pressure dependence in the gamma-ray flux above 3 MeV. Therefore, we suggest that the gamma-ray data above 3 MeV needs to be corrected for the local atmospheric pressure variations.

\end{abstract}

\keywords{NaI (Tl) detector; gamma-ray experiments;  barometric pressure correction}
\section{Introduction} \label{sec:intro}
Cosmic rays (CRs) interacting with the atmosphere produce various particles known as secondary cosmic rays (SCRs). Among these, neutron, proton, $\pi^+$, $\pi^0$, $\pi^-$ are generated first. Neutral pions decay into photons and charged pions decay into muons, which can later produce photons. These reactions result in the abundance of photon, neutron, and muon on the ground. Mostly, the neutron is used as a probe to study primary CRs (solar, galactic, or extragalactic origin), sun-earth connection, etc. Neutron monitors (NMs) have been used to collect the neutron data for more than seven decades. The meteorological effects on neutron flux due to pressure, temperature, humidity, wind, snow, atmospheric electric field have been studied extensively by \citet{dorman1972meteorologicheskie,dorman2004cosmic,belov1993,clem2000neutron,gerontidoupressure,platanoslong,thomas2017decadal}. The most prominent effect is due to the atmospheric pressure \citep{rochester1962cosmic,lindgren1962pressure,carmichael1968attenuation,chiba1976time}. Increase in the barometric pressure causes a decrease in the CR intensity, which has been reported since the 1920s and was interpreted as the effect due to absorption of CR by the atmosphere \citep{myssowsky1926unregelmassige}. Later, \citet{dorman1972meteorologicheskie} discussed the complex nature of barometric effect as a result of absorption, decay, and generation of SCRs. It is also known that the barometric effect varies with geomagnetic cutoff rigidity and altitude of the location of observation.  

To obtain useful data from NMs, correcting the NM data for the effect of barometric pressure is regarded as one of the most important corrections of raw data {\textit{(http://cosray.phys.uoa.gr/index.php/data/nm-barometric-coefficient)}}. Raw data from a neutron detector is corrected for pressure using experimental calculation of the barometric coefficient. The method to define the barometric coefficient of the different components of secondary cosmic rays has been studied by many researchers in the past \citep{carmichael1968attenuation,dorman1972meteorologicheskie}. The literature describes the effect of barometric pressure on the counting rate of NMs as
\begin{equation}\label{eq1}
  dN = -b.dp,
\end{equation}
where $dN$ is the rate of change in counts, $dp$ is the change in pressure and $b$ is the barometric coefficient. Integrating Eq. \ref{eq1} and assuming that there are $N_0$ counts for pressure $p_0$, the corrected counts ($N$) are calculated as
\begin{equation}\label{eq2}
N = N_0\exp[{-b(p-p_{0})}],
\end{equation}
where $p$ is the current atmospheric pressure. The parameters $N_0$ and $p_0$ are the reference values (usually, the average values) of counting rate and atmospheric pressure, respectively, over the specified time period. Eq. \ref{eq2} can be used for experimental calculation of the barometric coefficient, `$b$' by linear regression of $N$ and $p$ measurements for a specific period. Calculating `$b$' for an NM station has become much easier because of an online tool developed by the Athens Cosmic ray team as described by \citet{paschalis201310}, which uses the data available on NMDB. 

During the last 60 years, the factors responsible for the change in b have been investigated using the data from networks of the neutron as well as muon detectors. b depends on many factors such as latitude (rigidity), altitude, and time of observation. It is found to have 11-year solar cycle dependence \citep{platanoslong,gerontidoupressure}. b is different for different types of particles detected at the same location and time. The absolute value of b for neutrons is more than that for muons \citep{chilingarian20111140}. Based on ASEC muon data, \cite{chilingarian20111140} found that the absolute value of b is inversely proportional to the muon energy, indicating that b varies with the energy of the detected particles. Thus, we can say that the barometric pressure effect on neutron and muon components of SCR has been studied comprehensively in the past.
 
On the other hand, similar studies discussing the barometric pressure effect on the photon component are scarce. Being massless, photons are not expected to have a dependence on atmospheric pressure. Unlike neutron and muon, for which SCR is the only major source, photons originate from different sources. In addition to the SCRs, terrestrial radioactivity is also a major contributor to photon production. Terrestrial radioactivity produces photons up to 2.8 MeV, while SCR can produce photons of energies ranging from 100 keV to tens of MeV. Therefore, understanding the pressure dependence can be complex. To our knowledge, only a couple of papers have attempted to study the role of pressure in gamma-ray flux  \citep{bukata1962determination,chin1968barometric}. They reported an observation showing a correlation of photon intensity with the atmospheric pressure. Though the aerial radiation surveys are known to correct their data for atmospheric pressure, they are actually height corrections. Their interest lies in the terrestrial radioactivity emanating from the ground. Therefore, the height correction is required for the radiation coming from the ground (high pressure) to the aircraft (low pressure). The pressure correction for gamma-ray data measured on the ground is different than such height correction. \citet{chin1968barometric} evaluated b for different energy ranges and they found the energy dependence of b in the range 3.8 -- 183 MeV. They reported an increase in b with energy, although they did not discuss the physics behind this observation. It should be noted that their observations were made at a location with the geomagnetic cutoff rigidity, R$_c$ $\approx~$0.85 GV (49.9{$^\circ$}N, 97.2{$^\circ$}W) and an elevation of 236 m a.s.l.  Their set-up consisted of a NaI detector of size $9.5$" diameter and $8$" length (volume 238.76 inch$^3$ (606.45 cm$^3$)), which was completely surrounded by an efficient scintillating plastic anticoincidence shield. In their paper, they have reported the barometric effect for energies above 3.8 MeV, and have not discussed the effect on energies below 3.8 MeV. In fact, NaI (Tl) is considered to be very effective for gamma-ray spectroscopy up to 3 MeV, therefore, it is important to study the pressure aspect on energies below 3 MeV as well. In the present paper, we investigate barometric pressure effects on the gamma-ray data of energies between 150 keV and $\sim$10 MeV, collected using NaI of size $4\times4\times16$" (volume 256 inch$^3$ (650.24 cm$^3$)) located near the equator (R$_c$ $\approx$ 17.4 GV).  

\section{Data}
The gamma-ray data used in the present analysis is obtained from the NaI (Tl) detector located at Equatorial Geophysical Research Laboratory (EGRL), Tirunelveli (30 m a.s.l.; Geographic Coordinates: 8.71{$^\circ$}N, 77.76{$^\circ$}E). The detector is placed inside a temperature-controlled cabin constructed six feet above the ground. The detector is inside the lead shielding that covers the bottom and all four sides but not the top. The top of the cabin is covered by plywood. Digital pulse processing for pulse height analysis (DPP-PHA) is implemented to record an energy histogram each minute. The spectrum is calibrated using standard sources ($^{60}$Co and $^{137}$Cs) and distinct background radioactivity peaks such as $^{40}$K, $^{208}$Tl, $^{214}$Bi. The details of the experimental set-up are described in \citet{vichare20182555}. The data obtained from this set-up have been used successfully to study the diurnal variation in gamma rays \citep{datar2019causes} and the response during a cyclone event \citep{datarcyclone}. The atmospheric pressure data used in the present analysis is obtained from an in-house automatic weather station (AWS) at EGRL.
 
\section{$\gamma$-rays and pressure variations}
\subsection{Total $\gamma$-ray counts}
\begin{figure*}
\begin{center}$
\begin{array}{cc}
\includegraphics[width=8cm]{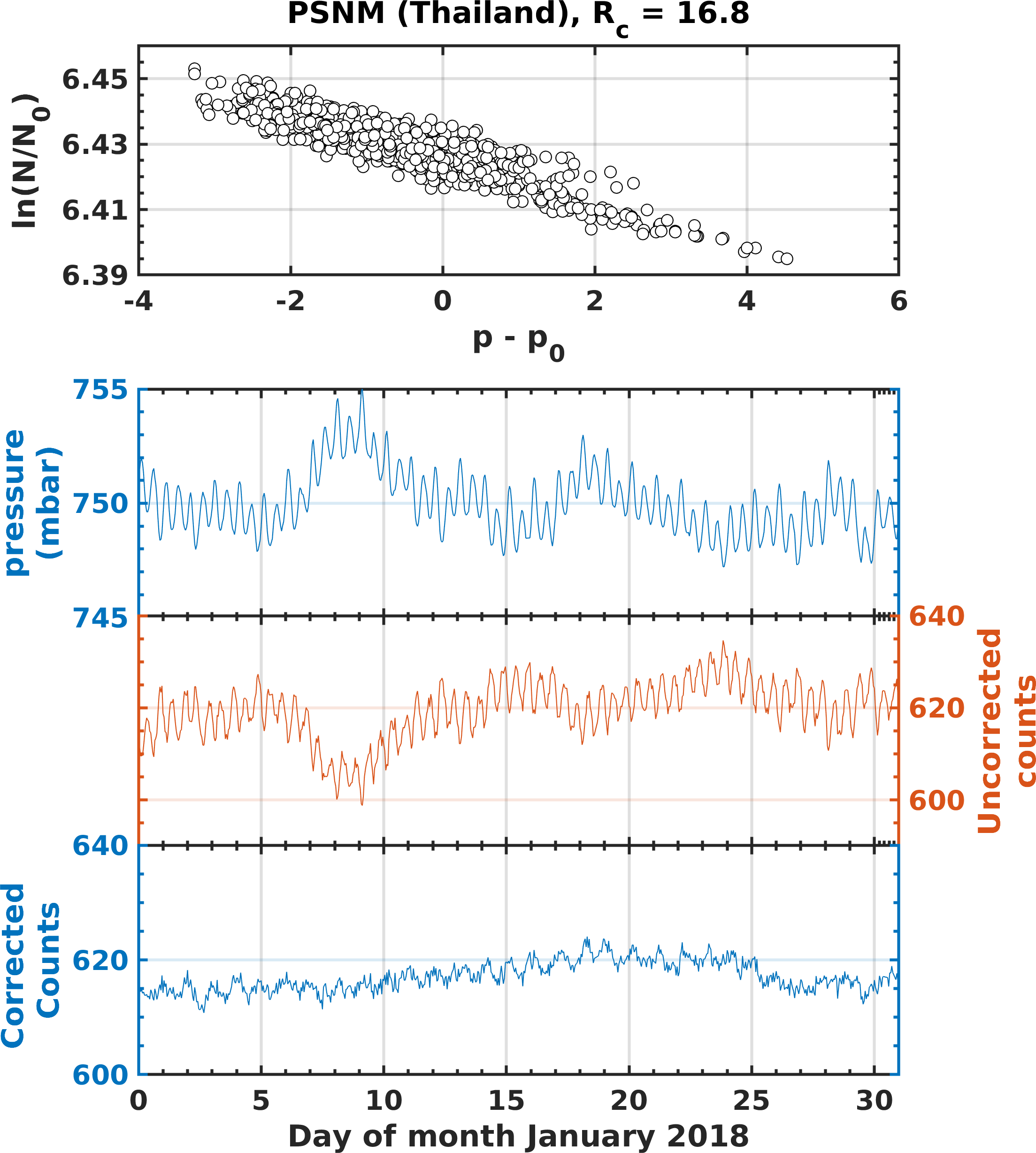} & \includegraphics[width=8cm]{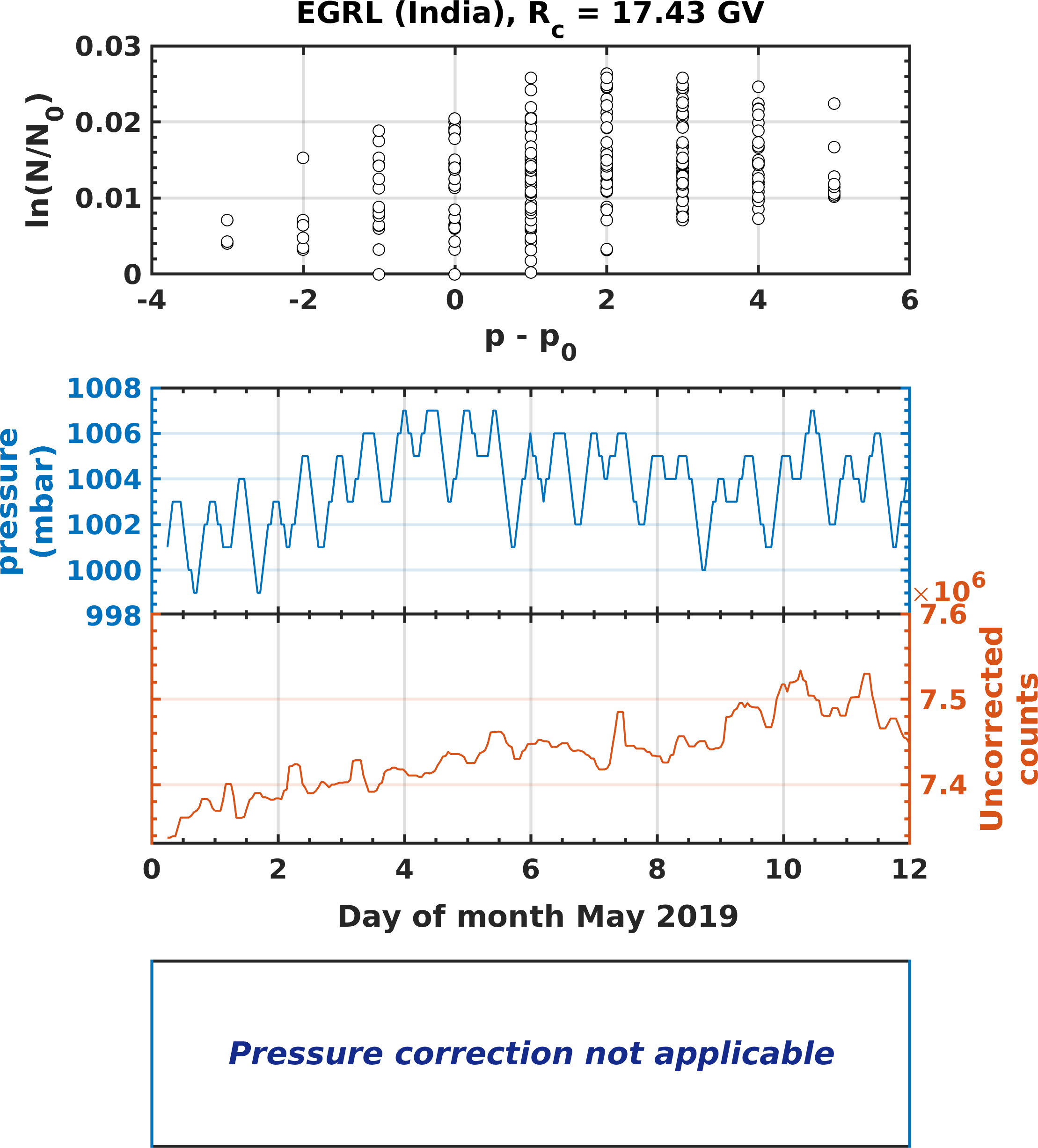}
\\
\end{array}$
\end{center}
\caption{\textit{Barometric pressure correction for NM data (Thailand station) and NaI (Tl) data}}
 \label{fig:psnm_egrl}
\end{figure*}

The dependence of the \textit{total} gamma-ray counts collected by the above set-up on the barometric pressure is examined in Figure \ref{fig:psnm_egrl}. Here, we have also presented the neutron data obtained from an NM station -- PSNM (Thailand), which has similar latitude and rigidity as Tirunelveli. We have randomly selected NM data from 1 to 31 January 2018. For NaI (Tl), data from 1 to 12 May 2019 is presented.  The x-axis denotes time in days. The top panels show the graphs of $ln(N/N_0)$ vs $p-p_0$. For NM, $p_0$ is the reference pressure value obtained from {\textit{http://www.nmdb.eu/station\_information/}}, which is a long-term average value. Similarly, for NaI data, $p_0$ is reference pressure taken as an average at Tirunelveli. $N_0$ is a base reference value of counts, which is taken as the minimum value of counts in the data. The plots on the left represent the NM data from PSNM, Thailand, while the plots on the right show gamma-ray data from NaI (Tl) located at EGRL, Tirunelveli. The second panels from the top show the barometric pressure recorded at the observation site. Third and bottom-most panels show uncorrected and corrected counts, respectively. It can be observed that the pressure and uncorrected counts of NM data are anti-correlated with each other over a long period ($\sim$10 days) and also on a diurnal scale. The uncorrected counts of NM data show anti-correlation with the semi-diurnal variations of the atmospheric pressure. On the other hand, no such dependence is observed in uncorrected counts from NaI (Tl) detector.

For PSNM, the scatter plot of $ln(N/N_0)$ vs $p-p_0$ is linear and the slope of a linear fit is taken as  b. Whereas for EGRL, the plot is totally scattered, indicating a lack of dependence and no reliable linear fitting is possible. Thus, it is evident from Figure \ref{fig:psnm_egrl} that the total counts collected by NaI (Tl) do not show any dependence on the atmospheric pressure, which is supported by the scatter plot depicted in the top panel.  It should be noted that these are the total counts detected by the NaI (Tl) detector integrated over all the energies. However, considering various sources of gamma rays (terrestrial radioactivity is the major source for gamma rays up to 2.8 MeV, while SCR is the dominant source above 3 MeV), it would be interesting to conduct a similar analysis for different energies, which is presented in the next subsection. 

\subsection{$\gamma$-ray counts of different energies}
\begin{figure*}[h!]
\begin{center}$
\begin{array}{c}
\includegraphics[width=16cm]{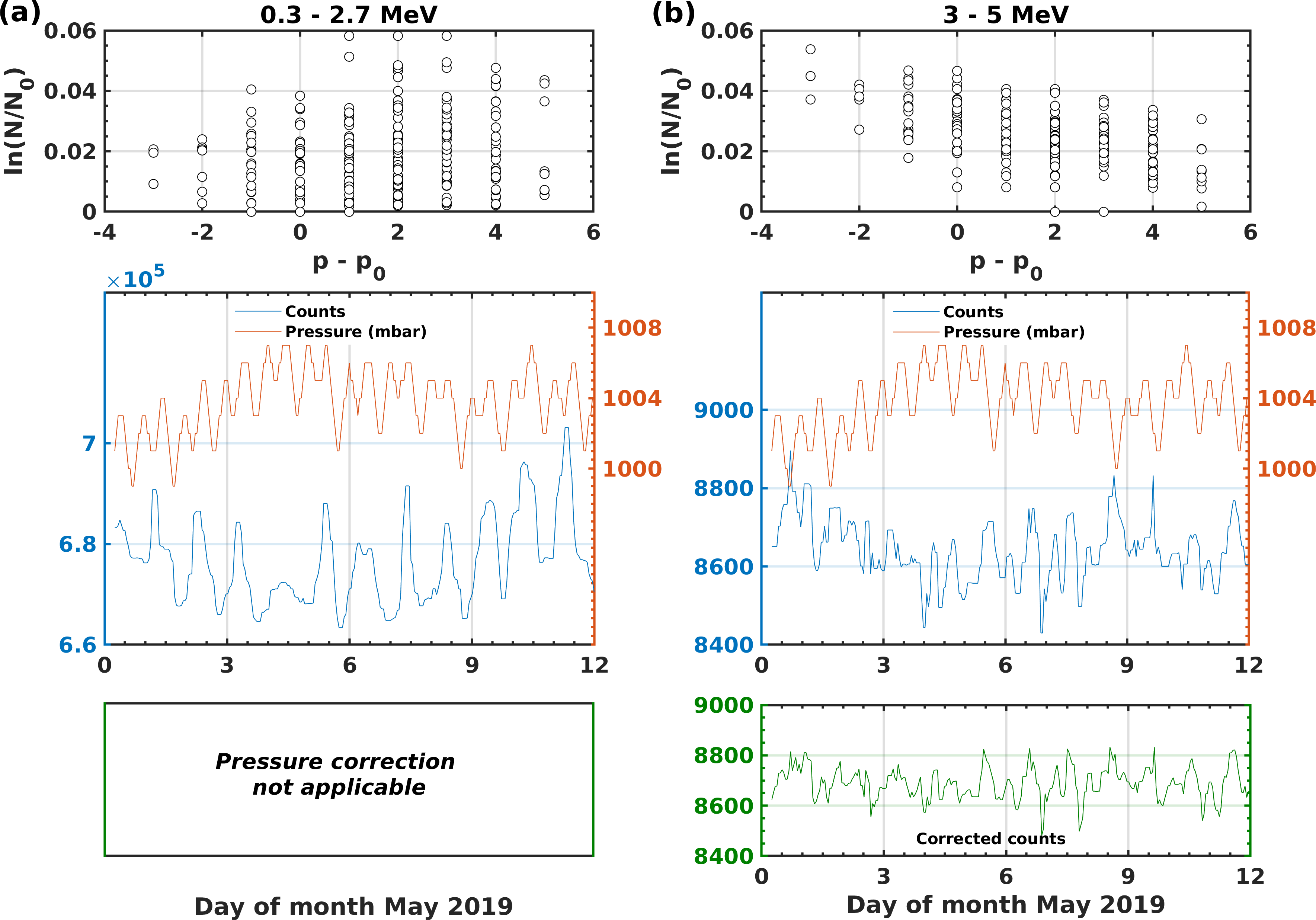} \\
\\
\includegraphics[width=16cm]{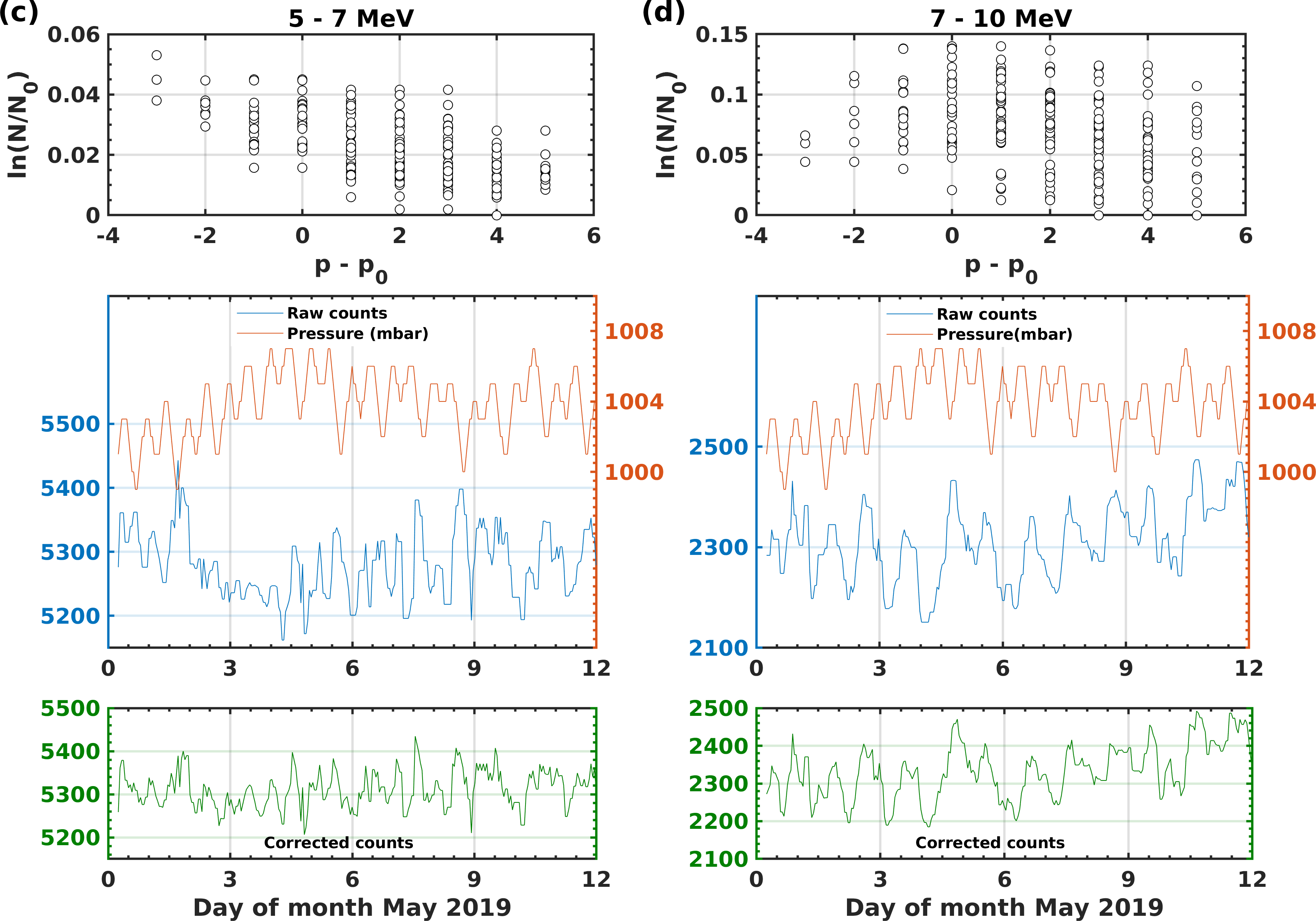}\\
\end{array}$
\end{center}
\caption{\textit{Barometric dependence of $\gamma$-ray counts of different energies during May 2019 (a) 0.3-2.7 MeV; (b) 3-5 MeV; (c) 5-7 MeV; (d) 7-10 MeV.}}
 \label{fig:diff}
\end{figure*}
The analysis for different energies of gamma flux during May 2019 is presented in Figure \ref{fig:diff} ((a) 0.3-2.7 MeV; (b) 3-5 MeV; (c) 5-7 MeV; (d) 7-10 MeV). The top panels in each subfigure show the graph of ln (N/N$_0$) on the y-axis and (p-p$_0$) in \textit{mbar} on the x-axis. As mentioned earlier, N$_0$ and p$_0$ are the reference values of counts and atmospheric pressure, respectively. The middle panel shows the atmospheric pressure (\textit{mbar}) and the uncorrected counts from NaI (Tl) detector within the specified energy range. The bottom panel shows the corrected counts wherever applicable, i.e., whenever good linear fit with negative slope is obtained, the correction is applied using the barometric correction formula (Eq. 2). The goodness of the fit is determined using the statistical parameters displayed in Table \ref{tab:statmay}. It can be observed from the scatter plots that the scatter is more for the lower energy band (0.3-2.7 MeV) than the other energy bands. Among the rest of the energies, 3-5 MeV and 5-7 MeV have less scatter and it is possible to fit a linear trend with a negative slope. For 7-10 MeV energy range as well, it is possible to fit a line with a negative slope. The slopes of the linear fits in the scatter plots represent b values and are shown in Table \ref{tab:statmay}. It can be noticed from Table \ref{tab:statmay} that the b value is negative above 3 MeV and small but positive for the energy range of 0.3-2.7 MeV. The b value computed for all energies as shown in Figure \ref{fig:psnm_egrl} is also a positive small value. The plots of pressure and uncorrected counts above 3 MeV seem to be anti-correlated on semi-diurnal as well as long time scales. But for energy range 0.3-2.7 MeV, such relationship is not clearly seen. The parameters such as Pearson correlation coefficient (CC), R$^2$, SSE, and RMSE enlisted in Table \ref{tab:statmay} describe the statistical significance of the linear fits and hence that of b values. The CC values for the parameters displayed in the scatter plots are in general good above 3 MeV. CC values are $\sim-0.6$ for energies between 3-7 MeV and dropped to $-0.294$ for 7-10 MeV range, though it is statistically significant (p $<$ 0.01). However, the CC for 0.3-2.7 MeV energy range is very small ($0.125$) indicating poor relation, and with p $>$ 0.01, it is not statistically significant as well. CCs of all energy ranges above 3 MeV have 99 \% confidence level (shown in bold in Table \ref{tab:statmay}). The SSE and RMSE values are small in general. It can be observed that for the energies above 3 MeV, the absolute value of b increases with energy from $0.3$ to $0.5$.  Further, based on the statistical significance of CC, we decided whether to apply the pressure correction or not. This information is shown in the last row of Table \ref{tab:statmay}. 
 
\begin{table*}[b!]
\begin{center}
\caption{Goodness-of-fit statistics for $\gamma$-ray counts of different energies (May 2019)}
\label{tab:statmay}
\begin{tabular}{|c|c|c|c|c|c|}
\hline
\textbf{Parameter} & \textbf{All energies} & \textbf{0.3-2.7 MeV} & \textbf{3-5 MeV} & \textbf{5-7 MeV} & \textbf{7-10 MeV}\\
\hline
\textbf{b (\%/mbar)} & 0.09174 & 0.08694 & -0.3044 & -0.3304 & -0.5237\\
\textbf{R$^2$\footnote{\textbf{R$^2$} indicates the success of the fit in explaining the variation of the data as it is the square of the correlation between the response values and the predicted response values. Value of R-square closer to 1 indicates good correlation.}} & 0.08507 & 0.01559 & 0.3646 & 0.3614 & 0.08649\\
\textbf{CC\footnote{\textbf{CC} is the Pearson correlation coefficient.}} & 0.2917 & 0.1248 & \textbf{-0.6038} & \textbf{-0.6012} & \textbf{-0.2941}\\
\textbf{SSE\footnote{\textbf{SSE} is `sum of squares due to error', which measures the total deviation of the response values from the fit to the response values. It is also called the summed square of residuals. A value closer to 0 indicates that the fit will be more useful for prediction.
}} & 0.00835 & 0.0440 & 0.0149 & 0.0178 & 0.267\\
\textbf{RMSE\footnote{\textbf{RMSE} is `root mean squared error' or the standard error of the regression. }} & 0.00547 & 0.0126 & 0.00731 & 0.00799 & 0.0309\\
\hline
\textbf{Correction} & NA & NA & Yes & Yes & Yes\\
\hline
\end{tabular}
\end{center}
\end{table*}

\begin{figure*}
\begin{center}
\includegraphics[width=15cm]{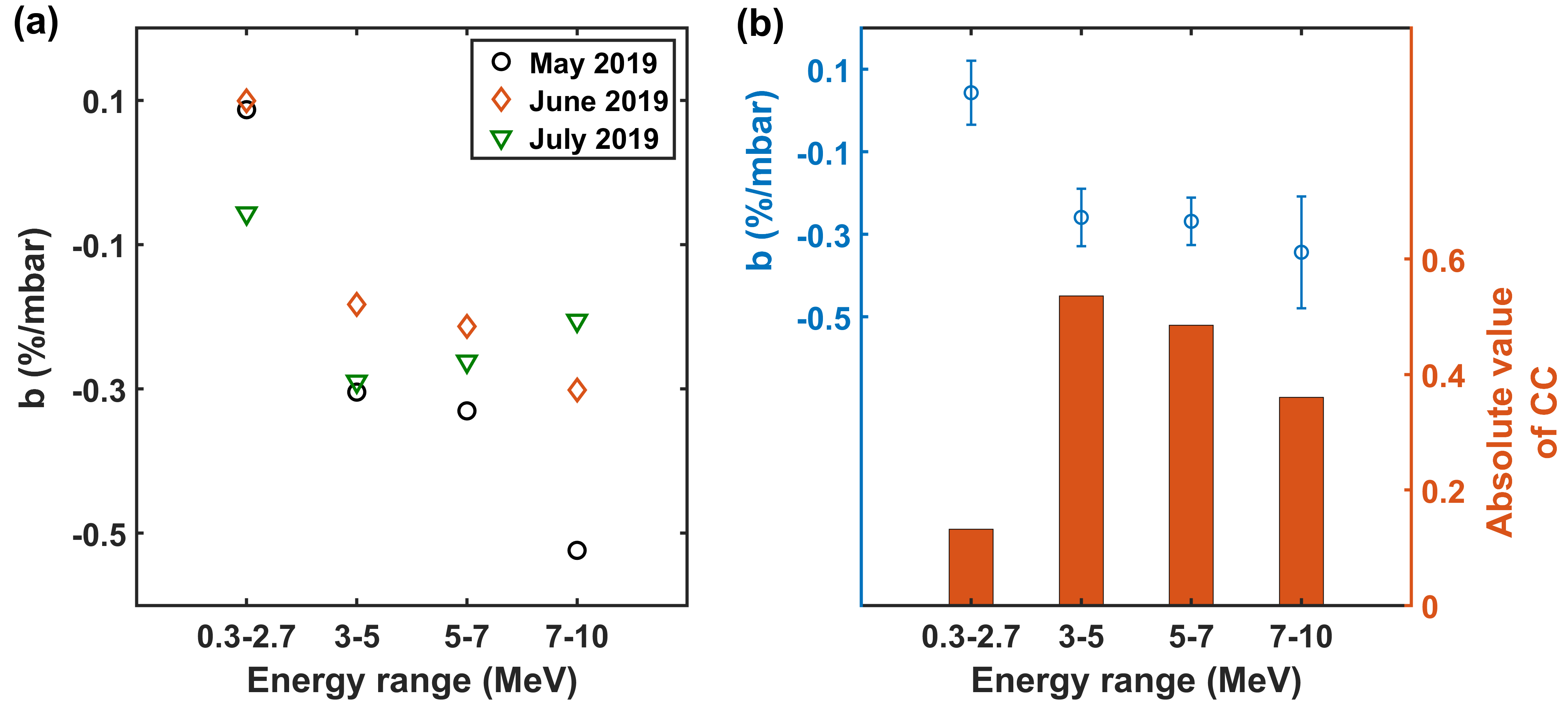} 
\end{center}
\caption{\textit{Summary and statistics of data from May, June, and July 2019}}
 \label{fig:stat}
\end{figure*}

We carried out similar analysis for two more data sets viz. six days from June 2019 and five days from July 2019, to confirm the results obtained above. The important observation in the present work is about the pressure dependence of gamma rays that varies with the energy of the gamma. The results from all these three data sets are plotted together in Figure \ref{fig:stat}. Figure \ref{fig:stat}(a) shows the variation of obtained barometric coefficient with energy. The mean b values in each energy range along with standard error (shown by vertical lines) are depicted in Figure \ref{fig:stat} (b). The significance of obtained b value is indicated by averaged CC values which are displayed by bar plots in Figure \ref{fig:stat} (b).  It can be observed that for 0.3-2.7 MeV energy range, the b values are very small and the average CC is also very small ($0.13$). This indicates that there is no pressure dependence for the energy range between 0.3-2.7 MeV.  For 3-5 MeV and 5-7 MeV energy ranges, the mean b is $\sim-0.26$ and $-0.27$, respectively with mean CC greater than $\sim$0.5, implying significant pressure dependence. For 7-10 MeV, mean b is $-0.34$ with mean CC of $0.36$. Larger error bars in this higher energy band can be due to the lower number of counts in this energy range. Thus, it is clear from Figure \ref{fig:stat} which is based on the data collected during different periods that there is no pressure dependence for the energy range 0.3-2.7 MeV, while clear dependence exists for the gamma of energies greater than 3 MeV and value of b increases with the energy above 3 MeV.

\section{Discussion}
The main objective of the present study is to investigate the pressure dependence of gamma-ray data collected by NaI(Tl) scintillation detector. NaI (Tl) detector is considered best for gamma spectroscopy up to 3 MeV. However, it can also detect gamma rays of energies higher than 3 MeV. Also, apart from gamma rays, it can detect other particles such as electrons, muons, neutrons as well, but the efficiency to detect these particles is very low. It may be important to discuss the contributions of other particles in the spectrum obtained from the NaI detector. As pointed out by \citet{Avakyan2013}, the ratio of efficiencies of a NaI (Tl) detector to detect gamma and neutrons varies from 5:1 to 8:1 in the energy range 3 MeV -- 10 MeV. This means that when a NaI (Tl) detects 50 to 80 \% of the incident gamma photons, it also detects 10 \% of the incident neutrons. Similarly, it might be having some efficiency to detect other particles like muons and electrons. However, as per EXPACS simulations, which compute the flux of various components of SCR \citep{sato2015analytical,sato2016analytical}, in the energy range of 113 keV --11.3 MeV, a total flux of gamma photons equals to 2.67E-01 cm$^{-2}$ s$^{-1}$. Similarly, for neutrons, it is 9.61E-03 cm$^{-2}$ s$^{-1}$ and  3.14E-05 cm$^{-2}$ s$^{-1}$ for muons ($\mu^{\pm}$). For electrons and positrons together, it is 1.04E-02 cm$^{-2}$ s$^{-1}$. Thus, with respect to gamma, the neutrons and $e^{\pm}$ have fluxes of 3.66 \% and 3.95 \%, respectively,  while muons have very less abundance of 0.012 \% in this energy range. Therefore, we can rule out the role of muons affecting our observations significantly. We cannot yet ignore the contribution of neutrons, electrons, and positrons ($\sim$7.61 \% flux as compared to gamma), despite the fact of lower efficiency of NaI (Tl) to detect these particles. We have to keep in mind that the pressure effect on neutrons is well studied, and in the given set-up, we cannot separate them from gamma. Positrons will not be registered even if they enter the detector, as they will get annihilated soon. Thus, in principle, we may have a contribution from neutron and electron in the counts measured by the NaI (Tl) detector. However, this is going to be small as illustrated by the following example for neutron. The efficiency of a NaI (Tl) to detect neutrons is $\sim$10 \%, and the average abundance w.r.t. gamma is 3.66 \%. That means if 100 gamma particles are incident on the detector, then 3.66 neutrons will also be incident on it. Out of those 100 gammas, $\sim$60-70 will be registered by NaI(Tl) and with 10 \% efficiency, only 0.366 neutrons will be detected. Thus, even though NaI can detect other particles, their contribution is not very significant and hence the particles detected by NaI are mostly gamma rays.

The present study finds that there is no correlation between the atmospheric pressure and the \textit{total} gamma-ray counts collected from all energies. However, the scenario differs when the gamma-ray flux of different energies are investigated separately. Besides the SCR, gamma rays observed on ground originate from other sources as well, such as terrestrial radioactivity, thunderstorms. Therefore, understanding gamma-ray dependence on atmospheric pressure is complex. 
The gamma particles with energies up to 2.7 MeV have a major contribution from the terrestrial radioactivity coming from the ground. While gamma as the SCR component produced in the CR cascade has a broad energy range from few keV to tens of MeV.  \cite{datar2019causes} have reported the presence of a diurnal variation in the gamma-ray counts with energy less than 2.7 MeV, which they attributed to the transport of radon in the air emanating from the ground to the atmospheric boundary layer. Therefore, the gamma radiation of energy $<$ 2.7 MeV originating from the ground do not pass through the overhead atmosphere and will not be affected by atmospheric pressure variations. Thus, our observation of the lack of pressure dependence of gamma flux in the energy range of 0.3-2.7 MeV is justified. Furthermore, as these energies have the highest counts associated with them in the energy histogram, it dominates the dependence relation when the total counts are considered (as seen in Figure \ref{fig:psnm_egrl}). Therefore, the total counts do not show the pressure dependence. 
It is evident from Figure \ref{fig:diff}, Figure \ref{fig:stat}  and Table \ref{tab:statmay} that the barometric pressure dependence exists for gamma rays above 3 MeV. However, gamma, i.e., photon being massless, is not expected to get affected by the atmospheric pressure variations. Therefore, the question is why the pressure dependence is observed in the present analysis for gamma with energy $>$ 3 MeV? It is true that once produced in the cascade reaction, the gamma does not interact with the atmospheric particles and hence does not depend on the pressure variations in the atmosphere. However, the pressure can affect the particles that generate photons. Photons are produced in the air shower during various interactions at different altitudes. Neutrons, $\pi^+$, $\pi^-$, $\pi^0$ are produced in the interactions of primary CRs with the atmospheric nuclei. $\pi^0$ directly produce photons, but charged pions decay into muons, which can produce photons after travelling some distance in the air. Thus, photons being \textit{tertiary} or \textit{quaternary} particles in this chain, their flux on the ground is a mixture of all those produced by different generation processes. This makes its dependence on the barometric pressure not so straightforward and maybe the reason for a not very good value of CC ($\sim-0.6$). Note that in case of neutrons and muons, the CC values with atmospheric pressure are $>~0.9$ \citep{chilingarian20111140}. The barometric coefficient found through the present work is around $-0.25$ to $-0.30$ \%/mbar for 3-10 MeV, which is similar to that of low-energy muons.

\section*{Conclusion}
Gamma rays, being massless, are not expected to be affected by pressure variations. However, atmospheric pressure affects the particles such as $e^{\pm}$, $\mu^{\pm}$, which produce gamma rays in the cascade of SCRs. Thus, the indirect pressure dependence of the gamma-ray flux is observed in gamma rays of energies above 3 MeV. However, gamma rays of energies below 2.7 MeV do not show such pressure dependence because of the dominance of the component of terrestrial origin, which does not get affected by atmospheric pressure variations. 

Present work suggests that the gamma-ray data collected on the ground for energies above 3 MeV need to be corrected for atmospheric pressure. Furthermore, the present study has been carried out at a location where the maximum pressure variation is only $\sim$7-8 mbar on an average, and yet we observe the pressure dependence in the gamma data. Thus, for the locations with stronger pressure variations, it is very important to consider the pressure corrections.

\section*{Acknowledgements}
The experimental set up at Tirunelveli is operated by the Indian Institute of Geomagnetism. This work is supported by the Department of Science and Technology, Government of India. Neutron monitor data was obtained from Neutron Monitor Database (NMDB) by courtesy of the Princess Sirindhorn Neutron Monitor Program. We acknowledge the corresponding authorities for making EXPACS software available online at \textit{http://phits.jaea.go.jp/expacs/}.

\section*{Author contributions statement}

All authors participated in designing the experiment. GD carried out the data analysis. GD and GV analysed the results and finalised the manuscript.

\section*{Additional information}
\textbf{Competing interests}: The authors declare no competing interests.
\bibliography{references}
\end{document}